# Hybrid Piezoelectric-Magnetic Neurons: A Proposal for Energy-Efficient Machine Learning

William Scott[1], Jonathan Jeffrey[1], Blake Heard[1], Dmitri Nikonov[2], Ian Young[2], Sasikanth Manipatruni[2], Azad Naeemi[1], Rouhollah Mousavi Iraei[1]

[1]Department of Electrical and Computer Engineering, Georgia Institute of Technology, Atlanta, GA, USA, iraei@gatech.edu
[2]Components Research Group, Intel Corporation, Hillsboro, OR, USA

**Abstract**

This paper proposes a spintronic neuron structure composed of a heterostructure of magnets and a piezoelectric with a magnetic tunnel junction (MTJ). The operation of the device is simulated using SPICE models. Simulation results illustrate that the energy dissipation of the proposed neuron compared to that of other spintronic neurons exhibits 70% improvement. Compared to CMOS neurons, the proposed neuron occupies a smaller footprint area and operates using less energy. Owing to its versatility and low-energy operation, the proposed neuron is a promising candidate to be adopted in artificial neural network (ANN) systems.

## I. Introduction

Deep learning enabled by developments in artificial neural networks (ANNs) has attracted special attention in recent years [32]. Cognitive learning researchers have used ANNs to simulate the natural learning process of the brain and improve the precision of speech recognition, the accuracy of pattern finding, and the reliability of self-driving cars [1, 4, 9, 11, 22]. Modern computer architectures struggle to emulate an ANN, even when processing on highly parallelized GPU architectures [7], [23]. To circumvent this challenge, researchers have turned to investigate how to integrate neural networks directly into hardware. Implementing ANNs as conventional CMOS hardware reduces the power consumption by three orders of magnitude [19]. Even with these improvements, CMOS neuron implementations are inefficient in energy consumption and die area, leading to increasing interest in beyond-CMOS devices for implementing neurons. Most notably, spin-based devices have been proposed as artificial neurons with simpler structure and lower energy consumption than their CMOS counterparts [17], [24]. These spintronic devices have shown to holistically mimic properties of neurons, providing advantages in circuit simplicity, adaptability, and energy efficiency [26]. Moreover, spintronic devices inherently offer non-volatile memory [5], [31]. ANNs need stored information for synaptic weights between communicating neurons; thus, having memory coupled with the circuit reduces energy dissipation and memory bandwidth, helping circumvent the von Neumann bottlenec

Several spin-based neurons are implemented using tunnel magnetoresistance (TMR) in magnetic tunnel junctions (MTJs) [35] coupled with various phenomena such as domain-wall (DW) motion [3], [27], spin transfer torque (STT) generated by lateral spin valves (LSVs) [17], [28], and spin-Hall effect (SHE) [27], [10]. While these devices are proven to mimic neural properties, some of their inherent drawbacks must be addressed. The slow switching speed of DW-based neurons prohibits them from being an ideal candidate for the fast implementation of a neuron. To provide non-reciprocity for the LSV neuron, the output magnet is preset by 90° reorientation to its saddle point of energy profile using the STT, generated by preset spin currents. However, the large required current yields substantial energy dissipation in the device. Recent studies on the magnetostriction-assisted all-spin logic (MA-ASL) device, a novel spin valve proposal made of a hybrid structure of magnets and piezoelectrics, have shown the reduction of switching energy by two orders of magnitude [12], [13]. The switching energy can be reduced in an MA-ASL device by employing a 90° magnetostrictive switching, experimentally demonstrated in [34] and shown to be more robust to thermal noise [16]. Using these recent advances, this paper proposes a spin-based neuron based on an MA-ASL device and an MTJ. The proposed structure integrates the advantages of previously proposed spintronic neurons with those of MA-ASL creating a structure that can be implemented into large-scale ANNs.

The rest of the paper is organized into three sections. Section II details the operation of the proposed device. Section III shows how the proposed device integrates into larger circuit schemes. Section IV analyzes the performance of the device. Finally, in section V we conclude the paper.

## II. Behind the MA-ASL Device

Figure 1 displays the structure of a simple MA-ASL device [13], made of the input magnet, Magnet 1, and the output magnet, Magnet 2, with each resting on top of a piezoelectric layer and connected by a metallic (Cu) interconnect. Both Magnet 1 and 2 have an easy axis along the ±**x** direction; meaning the energy profile of magnets is lower at these directions. To reorient the output magnet, a voltage is applied across the thickness of the piezoelectric layer to create an anisotropic strain along the **y** axis inside the piezoelectric layer transferred to Magnet 2. This strain couples to the magnetization through the magnetoelastic energy of the magnet; hence, the easy axis of the magnet rotates by 90°; thus, Magnet 2 reorients from the +**x** direction to the ±**y** direction.

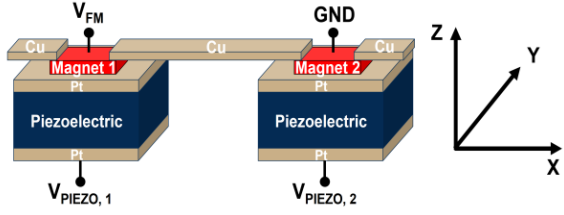

Figure 1. Schematic of the MA-ASL device.

Once V$_{PIEZO, 2}$ switches back to low, the easy axis will rotate back to the ±**x** direction, which briefly leave Magnet 2 at the meta-stable saddle point, making it equally likely to switch to the +**x** or -**x** directions. We break the symmetry by applying a spin-polarized current to Magnet 2 generated by an electrical current passing through Magnet 1; thus, the final magnetization is determined by the orientation of the spin current in either +**x** or -**x** directions. For instance, if a spin current with an orientation opposite to Magnet 1 is created by passing a current through the magnet to ground, the spin current will travel down the spin valve. This spin current applies an STT to Magnet 2, then Magnet 2 rotates to the -**x** direction after V$_{PIEZO2}$ goes low, making the device act as an inverter.

The physics behind the MA-ASL involves the magnetization dynamics of magnets, governed by the Landau-Lifshitz-Gilbert (LLG) equation,

$$\frac{d\vec{m}}{dt} = -\gamma\mu_0[\vec{m} \times \vec{H}_{eff}] + \alpha\left[\vec{m} \times \frac{d\vec{m}}{dt}\right] + \frac{\vec{I}_{s,\perp}}{qN_s}. \quad (1)$$

The magnetic orientation is represented by $\vec{m}$, and its change is related to $\vec{I}_{s,\perp}$, the perpendicular spin current, and $N_s$, the number of spins in the magnet. In (1), $\mu_0$, $\alpha$, and $\gamma$ represent the free space permeability, the Gilbert damping coefficient, and the gyromagnetic ratio, respectively [4], [29]. $\vec{H}_{eff}$ represents the net magnetic field, which is the sum of the uniaxial anisotropy field, $\vec{H}_U$, the demagnetization field, $\vec{H}_{demag}$, and the thermal noise field, $\vec{H}_{thermal}$ [5]. Anisotropy and magnetoelastic energy are intertwined; thus, $\vec{H}_U$ is created due to the changes in the magnetoelastic energy, $E_{ME}$ [24],

$$E_{ME} = \frac{-3}{2}\lambda Y\left[\left(m_x^2 - \frac{1}{3}\right)\epsilon_{xx} + \left(m_y^2 - \frac{1}{3}\right)\epsilon_{yy} + \left(m_z^2 - \frac{1}{3}\right)\epsilon_{zz}\right]. \quad (2)$$

The magnetostrictive coefficient and Young's modulus are represented by $\lambda$ and $Y$, respectively. The magnetization components are represented with $m_x$, $m_y$, and $m_z$, while the components of strain are represented with $\epsilon_{xx}$, $\epsilon_{yy}$, and $\epsilon_{zz}$ along the **x**, the **y**, and the **z** axes, respectively. The net applied strain to the magnet, $\epsilon_{xx} - \epsilon_{yy}$, is derived using

$$\epsilon_{xx} = \epsilon_0 + d_{31}\frac{V_{PIEZO,2}}{t}, \quad (3)$$

$$\epsilon_{yy} = \epsilon_0 + d_{32}\frac{V_{PIEZO,2}}{t}. \quad (4)$$

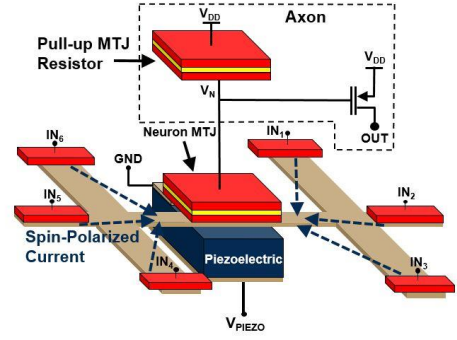

Figure 2. Proposed MA-ASL neuron, shown with six inputs. The net spin current in the interconnect applies STT to the free layer of the neuron MTJ in timing with the piezo clock, switching the orientation of the neuron output.

In these equations, $d_{31}$ and $d_{32}$ are piezoelectric constants [13], $V_{PIEZO,2}$ is the voltage applied across the piezoelectric layer, and $t$ is the piezoelectric thickness.

To simulate the operation of the MA-ASL device, magnets, metallic channels, and piezoelectric layer are modeled as building blocks that are solved self-consistently using SPICE [12], [13]. These models account for the change in the magnetoelastic energy of magnets as an equivalent anisotropy field [12], [13]. The magnetization dynamics and the spin current transport in metallic channels are accounted using the SPICE models developed by Bonhomme et al. in [5]. The magnetization dynamics is calibrated with numerical methods implemented by MATLAB in [5]. The current transport models are calibrated with experimental results [20].

## III. Spin Neuron Proposal

### A. Neuron Functionality

The proposed neuron, shown in Figure 2, is a modified MA-ASL structure whose output magnet is the free magnetic layer of an MTJ. The input voltages, shown for six inputs (IN$_1$–IN$_6$) in Figure 2 as an example, produce charge currents that flow through the corresponding input magnets and become spin-polarized at the interfaces with the metallic channel. These spin-polarized currents combine below the output magnet according to the sum,

$$I_{s,out} = \sum_j I_{s,j} = \sum_j \eta_j e^{\frac{-L_j}{L_{SRL,j}}} I_{c,j}, \quad (5)$$

where $I_{s,out}$ is the spin current injected into the output magnet, $I_{s,j}$'s are the spin current contributions from each magnet $j$, and $I_{c,j}$'s are the input charge currents [23]. The distance between each input magnet and the output magnet is represented by $L_j$. The spin polarization at the interface of each magnet and channel is represented by $\eta_j$. The spin relaxation length, $L_{SRL,j}$, is affected by the grain boundary and sidewall scattering due to size effects and material properties of the metallic channel [12].

The net injected spin current, $I_{s,out}$, applies an STT to the output magnet. If strong enough, the STT will rotate the output magnetization, $\hat{m}_{out}$. The output magnet is in contact with an MgO layer that separates it from a magnet fixed in the +**x** direction, forming a three-layer MTJ. As the output magnetization changes, the resistance across the MTJ also changes, following the equation,

$$R_{MTJ} = \frac{1+P}{G_P(1+P\hat{m}_{out,X})}, \quad (6)$$

where $R_{MTJ}$ is the resistance of the MTJ, $\hat{m}_{out,X}$ is the x-component of $\hat{m}_{out}$, and $G_P$ is the conductance of the MTJ in its low-resistance state, the +**x** direction [23]. The polarization factor, $P$, is

$$P = \frac{G_P - G_{AP}}{G_P + G_{AP}} = \frac{TMR}{TMR+2}, \quad (7)$$

where $G_{AP}$ is the conductance of the MTJ in its high-resistance antiparallel state, the -**x** direction [23]. As shown in Figure 2, the change in the resistance of the MTJ is sensed by connecting the structure to a pull-up resistor connected to $V_{DD}$; then, the voltage above the output neuron follows

$$V_N = \frac{R_{MTJ}}{R_{MTJ}+R_{Pull-up}} V_{DD}, \quad (8)$$

where $R_{Pull-up}$ is the resistance of the pull-up resistor, implemented with an MTJ with two fixed magnetic layers. The voltage, $V_N$, is amplified by a PMOS transistor, forming the axon where the neuron's output can be transferred to other neurons.

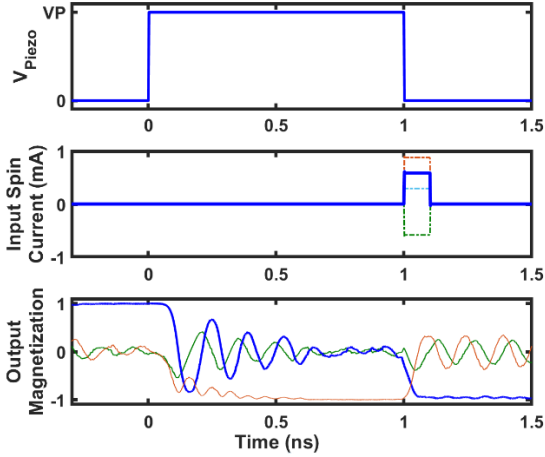

Figure 3. Transient response of the MA-ASL device. In the first phase of operation, $V_{PIEZO}$ turns on for 1 ns as shown in the first graph. The second graph illustrates the second phase of operation, in which STT is applied to the output magnet through the injected net spin current (in blue) from three input magnets (shown with dotted lines), applied after $V_{PIEZO}$ turns off. The third graph shows the magnetization of the output magnet (**x**, **y**, and **z** axes shown in blue, red, and green respectively), and how it is affected by $V_{PIEZO}$ and the spin currents.

### B. Transient Response of the Neuron

The transient response of the magnetization is shown in Figure 3 for a neuron with three inputs. In the first phase of device operation, $V_{PIEZO}$ is pulsed high for a duration of 1 ns, rotating $\hat{m}_{out}$ to the +**y** or the -**y** direction. When $V_{PIEZO}$ turns off, $\hat{m}_{out}$ will be placed at the saddle-point of the energy profile. In the second phase of operation, 10x shorter than the first phase, the input voltages are pulsed for 0.1 ns, applying an STT that tips $\hat{m}_{out}$ toward +**x** or -**x**. The delay of the final switching is inversely proportional to the magnitude of the net spin current, $I_{s,out}$. Compared to an STT-only realignment, this magnetostriction-assisted re-alignment of $\hat{m}_{out}$ onto the axis requires two orders of magnitude lower energy dissipation.

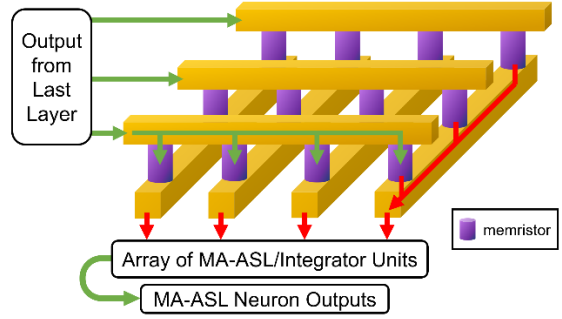

Figure 4. Memristive cross-bar network. The cross-bar array sums together the input currents, abbreviating the number of magnets needed for the output neurons.

**Table 1: Performance Comparison of MA-ASL Neuron against its CMOS and Spintronic Counterparts**

| NEURON DEVICE | Digital CMOS [21] | Analog CMOS [30] | Spintronic [29] | MA-ASL Neuron |
|---|---|---|---|---|
| DELAY | 10 ns | 10 ns | 1 ns | 1.1 ns |
| ENERGY | 832.6 fJ | 700 fJ | 0.81 fJ | 0.25 fJ |

### C. Integration into Neural Network

To connect the proposed device into a neural network with machine learning capabilities, we must first show how it mimics a neuron. In Figure 2, the axon of the neuron uses the voltage from the output MTJ as the gate voltage for a PMOS transistor, creating a charge current output. For the synapses, additional circuitry would be required to correctly weight the input current. One proposed method is with a memristive crossbar network, as shown in Figure 4. This structure places memristors between input and output lines to weight the charge current being passed among neurons [15]. In this setup, each output from the previous layer of neurons connects as an input to the crossbar network, which applies synaptic weights and outputs to the next layer of neurons.

## IV. Benchmarking Against Competing Technologies

As Figure 3 illustrates, the delay of the MA-ASL neuron is about 1.1 ns, slightly larger than that of the spintronic neuron presented in [27], which claims 1 ns. However, Table 1 demonstrates that the MA-ASL neuron demonstrates 70% improvement in terms of energy over the spintronic neuron [5]; the spintronic neuron uses STT to reorient magnets, while the MA-ASL neuron utilizes a combination of STT and magnetostrictive switching, which results in lower overall energy dissipation. When compared with both analog and digital CMOS neurons, the MA-ASL neuron has advantages in terms of energy consumption and overall chip area. These advantages are due to a more efficient implementation of a spintronic neuron that requires a lower device count. CMOS neurons require shift registers, sense amplifiers, DRAM, and SRAM, which all require large numbers of transistors [7], whereas spintronic neurons require one MTJ and one magnet for each input, using two orders of magnitude less area than CMOS [29] and three orders of magnitude less energy. These improvements in area and energy consumption enable the proposed device to excel in mimicking a neural network, providing competition to CMOS and other spintronic neural networks in Boolean and non-Boolean computations.

## V. Future Work

The efficiency of the proposed neuron in learning tasks can be tested through network-scale simulations. Moreover, beyond characterizing the transient response of a single MA-ASL neuron, a neural network architecture of multiple MA-ASL neurons must be investigated further. A prime candidate for a neural network implementation is a memristive crossbar network due to the inherent learning capabilities of memristors and the lower device count for the structure, because of elimination of circuitry required for backpropagation [15]. As a result, area and power consumption for a neural network will be reduced. The research on MA-ASL neural network topologies may lead to the implementation of network hierarchies usable for processor design or convolutional networks for deep learning [2], [21].

## VI. Conclusion

We proposed a spintronic neuron based on the MA-ASL device and the MTJ. The performance of the neuron is benchmarked against its CMOS and spintronic counterparts in terms of area, delay, and energy dissipation. The MA-ASL neuron operates with less than half the energy compared to its spintronic counterparts by employing magnetostrictive switching along with STT switching. Magnetostrictive switching is expected to further enhance the robustness of the operation of neuron to thermal noise as well. The operation of the device was simulated using SPICE models and the physics behind the operation of the device is well understood.